\documentclass{aa}  
\usepackage{graphicx}
\usepackage{natbib}
\usepackage{txfonts}
\def\lapp{\ifmmode\stackrel{<}{_{\sim}}\else$\stackrel{<}{_{\sim}}$\fi}
\def\gapp{\ifmmode\stackrel{>}{_{\sim}}\else$\stackrel{<}{_{\sim}}$\fi}

\begin{document}
   \title{The binary pulsar PSR\,J1811$-$1736: evidence of a low amplitude
     supernova kick.}

   \author{A.Corongiu
          \inst{1}\fnmsep\inst{2}\fnmsep\inst{3}
          \and
          M.Kramer\inst{3}
          \and
          B.W.Stappers\inst{4}
          \and
          A.G.Lyne\inst{3}
          \and
          A.Jessner\inst{5}
          \and
          A.Possenti\inst{2}
          \and
          N.D'Amico\inst{1}\fnmsep\inst{2}
          \and
          O.L{\"o}hmer\inst{5}
          }

   \offprints{A.\,Corongiu {\email{corongiu@ca.astro.it}}}

   \institute{Universit\`{a} degli Studi di Cagliari, Dip. di Fisica, S.P. Monserrato-Sestu km 0,700
              I-09042 Monserrato, Italy
         \and 
              INAF - Osservatorio Astronomico di Cagliari. Loc. Poggio dei Pini,
              Strada 54, I-09012 Capoterra, Italy
         \and
              University of Manchester, Jodrell Bank Observatory, Jodrell Bank, Macclesfield,
              Cheshire, SK11 9DL
         \and
              Stichting ASTRON, Postbus 2, 7990 AA Dwingeloo, The Netherlands.
         \and
              MPI f{\"u}r Radioastronomie, Auf dem H{\"u}gel 69, 53121 Bonn, Germany
             }

   \date{Received October 20, 2005; Accepted November 1,2006}

 
 \abstract
   {}
   {The binary pulsar PSR\,J1811-1736 has been identified, since its
   discovery, as a member of a double neutron star system.
   Observations of such binary pulsars allow the measurement of
   general relativistic effects, which in turn lead to
   information about the orbiting objects, to their binary evolution
   and, in a few cases, to tests of theories of gravity.}
    {{Regular timing observations have since 2000 been carried out with
    three of the largest European radio telescopes involved in pulsar
    research. Pulse times of arrival were determined by convolving the
    observed profiles with standard templates, and were fitted to a
    model that takes into account general relativistic effects in
    binary systems. The prospects of continued observations were
    studied with simulated timing data. Pulse scattering times were
    measured using dedicated observations at 1.4\,GHz and at
    3.1\,GHz, and the corresponding spectral index has also been
    determined.
    The possibility of detecting the yet unseen companion as
    a  radio pulsar was investigated as a function of pulse period,
    observing frequency and flux density of the source. A study of the
    natal kick received by the younger neutron star at birth was
    performed considering the total energy and total angular momentum
    for a two body system.}}
   {We present an up to date and improved timing
   solution for the binary pulsar PSR\,J1811-1736. One post-Keplerian
   parameter, the relativistic periastron advance, is measured and
   leads to the determination of the total mass of this binary system.
   Measured and derived parameters strongly support the double neutron
   star scenario for this system. The pulse profile at 1.4\,GHz is
   heavily broadened by interstellar scattering, limiting the timing
   precision achievable at this frequency. We show that a better
   precision can be obtained with observations at  higher
   frequencies. This would allow one to measure a second Post-Keplerian
   parameter within a few years. We find that interstellar
   scattering is unlikely to be the reason for the  continued failure
   to detect radio pulsations from the companion of PSR\,J1811-1736.
   The probability distribution that we derive for the amplitude  of
   the kick imparted on the companion neutron star at its birth
   indicates that the kick has been of low amplitude.}
   {}
   \keywords{Pulsar: General, PSR\,J1811-1736}

   \titlerunning{The binary pulsar PSR\,J1811$-$1736.}
   \maketitle
%

\section{Introduction}

The pulsar J1811$-$1736, discovered during observations of the Parkes
Multibeam Pulsar Survey\citep{mlc+01},  has a spin period of 104 ms
and is member of a 18.8-d, highly eccentric binary system
\citep{lcm+00} with an as yet undetected companion \citep{mig00}. The
characteristic age and the estimated surface magnetic field strength
indicate that the pulsar is mildly recycled. In such a system, it
is expected that the observed pulsar is born first in a supernovae
(SN) explosion before it undergoes mass accretion from a high-mass
non-degenerate binary  companion. Parameters which have been
measured and derived from the best-fit timing solution indicate that
the companion is quite massive. All these elements suggest that the
companion is also a neutron star (e.g. \citealt{bv91}).

The conclusion that PSR\,J1811$-$1736 is a member of the small sample
of known double neutron star (DNS) systems was already reached by
\citet{lcm+00}. Their conclusion was supported by a constraint on the
total system mass, assuming that the observed advance of periastron
was totally due to relativistic effects. However, the rather long
period of the pulsar, combined with the effects of interstellar
scattering, resulting in significant broadening of the pulse profile at
1.4~GHz, as well as the short data span available to \citet{lcm+00},
limited the timing precision and hence the accuracy of the total mass
measurement.

Among all DNS systems, PSR\,J1811-1736 has by far the longest spin
period, the longest orbital period and the highest eccentricity. This
may suggest that the evolution of this DNS has been different, at
least in part, from all other DNS systems. On the other hand,
PSR\,J1811-1736 well fits the spin period versus eccentricity relation
for DNS systems (\citealt{mlc+05}, \citealt{fkl+05}). This relation
can be simply explained in terms of the different lengths of time the
pulsar underwent accretion, which in turn is related to the mass of
the companion star before the SN explosion. Moreover, numerical
simulations \citep{dpp05} show that the spin period versus
eccentricity relation is recovered assuming that the second born
neutron star received a low velocity kick at its birth.

In this paper we report on new timing observations which significantly
improve on the previously published results. We present a study
of the observed interstellar scattering and consider its consequences
on the detectability of radio pulses from the companion. Finally, we
investigate the likely kick velocity imparted to the second-born
neutron star during its birth in the system's second SN.
Observations were carried out as part of a coordinated effort using
three of the largest steerable radio telescopes in the world for
pulsar timing observations, i.e. the 100-m radio telescope at
Effelsberg, the 94-m equivalent Westerbork Synthesis Radio Telescope
(WSRT) and the 76-m Lovell telescope at Jodrell Bank telescope. This
paper is the first in a series detailing results of these efforts in
establishing a {\em European Pulsar Timing Array} (EPTA).

\section{Observations}

The binary pulsar J1811$-$1736 is one of the binary pulsars regularly
observed by the EPTA. The aims and objectives of this European
collaboration include the detection of a cosmological gravitational
wave background and the project will be described in detail in a
forthcoming publication. Here we summarize the observing systems used
while further details can be found in the references below.

\subsection{Effelsberg timing}

We made regular timing observations of PSR\,J1811$-$1736 since October
1999 using the 100-m radio telescope of the Max Planck Institut f\"ur
Radioastronomie in Effelsberg near Bonn. The typical observing rate
was of 1 observation every two months. An overall root-mean-square
(RMS) of 538 $\mu s$ is achieved after applying the final timing
model.  The data were obtained with a 1.3$-$1.7\,GHz tunable HEMT
receiver installed in the primary focus of the telescope. The noise
temperature of this system is 25~K, resulting in a system temperature
from 30~to 50~K on cold sky depending on elevation.  The antenna gain
at these frequencies is 1.5~K~Jy$^{-1}$.

An intermediate frequency (IF) centred on 150 MHz for left-hand (LHC)
and right-hand (RHC) circularly polarised signals was obtained after
down-conversion from a central RF frequency of usually 1410 MHz.  The
signals received from the telescope were acquired and processed with
the Effelsberg-Berkeley Pulsar Processor (EBPP) which removes the
dispersive effects of the interstellar medium on-line using ``coherent
de-dispersion'' \citep{hr75}. Before entering the EBPP, the two LHC
and RHC signals of the IF are converted to an internal IF of 440
MHz. A maximum bandwidth of $2\times32\times0.7$~MHz~=~$2\times32$~MHz
was available for the chosen observing frequency and DM of the
pulsar. It was split into four portions for each of the two circular
polarisations, which were mixed down to baseband. Each portion was
then sub-divided into eight narrow channels via a set of digital
filters \citep{bdz+97}. The outputs of each channel were fed into
de-disperser boards for coherent on-line de-dispersion. In total 64
output signals were detected and integrated in phase with the
predicted topocentric pulse period.

A pulse time-of-arrival (TOA) was calculated for each average profile
obtained during a 5-10 min observation. During this process, the
observed time-stamped profile was compared to a synthetic template,
which was constructed out of 5 Gaussian components fitted to a
high signal-to-noise standard profile (see Kramer et al.,
\citeyear{kxl+98,kll+99}).  This template matching was done by a
least-squares fitting of the Fourier-transformed data \citep{tay92}. 
Using the measured time delay between the actual
profile and the template, the accurate time stamp of the data provided
by a local H-MASER and corrected off-line to UTC(NIST) using recorded
information from the satellites of the Global Positioning System
(GPS), the final TOA was obtained.  The uncertainty of each TOA was
estimated using a method described by \citet{dr83} and \citet{lan99}.

\subsection{Jodrell Bank timing}

Observations of PSR\,J1811-1736 were made regularly using the 76-m
Lovell telescope at Jodrell Bank since its discovery in 1997
\citep{lcm+00}.  The typical observing rate was of about 2
observations each week, with an overall RMS of 1300\,$\mu$s after
applying the final timing model. A cryogenic receiver at 1404\,MHz was
used, and both LHC and RHC signals were observed using a
$2\times32\times1.0$-MHz filter bank at 1404\,MHz. After detection,
the signals from the two polarizations were filtered, digitised at
appropriate sampling intervals, incoherently de-dispersed in hardware
before being folded on-line with the topocentric pulse period and
written to disk.  Each integration was typically of 1-3 minutes
duration; 6 or 12 such integrations constituted a typical
observation. Off-line, the profiles were added in polarisation pairs
before being summed to produce a single total-intensity profile.  A
standard pulse template was fitted to the observed profiles at each
frequency to determine the pulse times-of-arrival (TOAs). Details of
the observing system and the data reduction scheme can be found
elsewhere (e.g.~\citealt{hlk+04}).

\subsection{Westerbork timing}

Observations of PSR\,J1811$-$1736 were carried out approximately
monthly since 1999 August 1st, obtaining an overall timing RMS of
659~$\mu$s after applying the final timing model, at a central
frequency of 1380 MHz and a bandwidth of 80 MHz. The two linear
polarisations from all 14 telescopes were added together in phase by
taking account of the relative geometrical and instrumental phase
delays between them and then passed to the PuMa pulsar backend
\citep{vkv+02}. The data were obtained with the L-band receiver
installed in the primary focus of the telescopes. The noise
temperature of this system is 25~K, resulting in a system temperature
from 30~to 50~K on cold sky depending on elevation.  The antenna gain
at these frequencies is 1.2~K~Jy$^{-1}$. PuMa was used in its digital
filterbank mode whereby the Nyquist sampled signals are Fourier
transformed and the polarisations combined to produce total intensity
(Stokes I) spectra with a total of 512 channels. These spectra were
summed online to give a final sampling time of 409.6 $\mu$s and
recorded to hard disk.  These spectra were subsequently dedispersed
and folded with the topocentric period off-line to form integrations
of a few minutes.  TOAs were calculated for each profile following a
scheme similar to that outlined above for Effelsberg data, except a
high signal-to-noise standard profile was used instead of Gaussian
components.  In the future, EPTA timing will employ an identical
synthetic template for all telescopes.

\section{Data analysis}

The TOAs, corrected to UTC(NIST) via GPS and weighted by their
individual uncertainties determined in the fitting process, were
analysed with the {\tt TEMPO} software package \citep{tw89}, using the
DE405 ephemeris of the Jet Propulsion Laboratory \citep{sta90}. {\tt
TEMPO} minimizes the sum of weighted squared {\it timing residuals},
i.e.~the difference between observed and model TOAs, yielding a set of
improved pulsar parameters and post-fit timing residuals. A summary
of the basic characteristics of each dataset is shown in Table 1.

   \begin{table}
    \centering
      \caption[]{Data sets' characteristics}
         \label{TabData} 
         \begin{tabular}{llll}
            \hline
            \noalign{\smallskip}

& Jodrell Bank & Effelsberg & Westerbork\\
            \noalign{\smallskip}
            \hline
            \noalign{\smallskip}
N.~of~ToAs   & 348 & 74 & 213\\
Time~Span & 50842-53624 & 51490-53624 & 51391-53546 \\
R.M.S. & 1300 & 538  & 659\\

            \noalign{\smallskip}
            \hline
            \noalign{\smallskip}
         \end{tabular}
\end{table}

Although the templates used for the three telescope data differed,
resulting offsets were absorbed in a global least-squares
fit. Remaining uncertainties were smaller than the typical measurement
accuracy of the Jodrell Bank timing data of about 9 $\mu$s.

Before all TOAs were combined, preliminary fits were performed on each
dataset alone, in order to study possible systematic differences between
the datasets. We applied a small quadrature addition 
and a scaling factor to the
uncertainties to obtain the expected value of a reduced $\chi^2=1$
for each dataset. 
The final joint fit to all TOAs resulted in a $\chi^{2}$ value of
unity, avoiding the need to add further systematic uncertainties.

Table 2 summarizes all observed timing and some derived
parameters. For the observed parameters the quoted errors are twice
the nominal TEMPO errors.  For the derived parameters, the given
uncertainties are computed accordingly.

The joint fit allowed us to determine the spin, positional and
Keplerian orbital parameters plus one post-Keplerian parameter with a
precision better than the best determination from a single data set
alone. However, the high degree of interstellar scattering (Fig. 3)
means that further post-Keplerian parameters will be difficult to
measure with continued observations at this frequency. We will discuss
the future prospects for higher frequency observations in \S~5.

   \begin{table}
      \caption[]{Timing and derived parameters}
         \label{TabTim} 
         \begin{tabular}{ll}
            \hline
            \noalign{\smallskip}
Timing parameters & Joint~data~sets\\
            \noalign{\smallskip}
            \hline
            \noalign{\smallskip}
RA~(J2000, hh:mm:ss)     & 18:11:55.034(3)\\
DECL~(J2000, deg:mm:ss)  & -17:36:37.7(4)\\
Period, $P$~(s)      & 0.1041819547968(4) \\
Period derivative, $\dot{P}$~($10^{-19}$ s~s$^{-1}$)    & 9.01(5)\\
Dispersion Measure, DM~(pc~cm$^{-3}$)   &  476(5)\\
Projected semi-major axis$^{a}$, $a~\sin~i$~(s) & 34.7827(5)\\
Eccentricity, $e$       & 0.828011(9)\\
Epoch of periastron, $T_{0}$~(MJD)   & 50875.02452(3)\\
Orbital period, $P_{B}$~(d)     & 18.7791691(4)\\
Longitude of periastron, $\omega$~(deg)  & 127.6577(11)\\
Advance of periastron, $\dot{\omega}$~(deg~yr$^{-1}$) & 0.0090(2)\\
Flux density at 3100\,MHz, $S_{3100}$ (mJy) & 0.34(7) \\
&\\
Time~Span~(MJD) & 50842-53624\\
N.~of~ToAs   & 635\\
RMS~($\mu s$)    & 851.173\\
            \noalign{\smallskip}
            \hline
            \noalign{\smallskip}
Derived parameters$^{b}$ & \\
            \noalign{\smallskip}
            \noalign{\smallskip}
Characteristic~Age, $\tau_{\rm c}$ ($10^{9}~yr$)  & 1.83\\
Surface magnetic field, $B_{0}$~($10^{9}$ G)  & 9.80 \\
Total~Mass~$M_{\rm TOT}$ ($M_{\odot}$) & 2.57(10)\\
Mass~Function~$f(M_{\rm C})$ ($M_{\odot}$) & 0.128121(5)\\
Orbital separation $A$ (ls) & 94.4(6) \\
Minimum~companion~mass~$M_{\rm C,min}$ ($M_{\odot}$) & 0.93\\

            \noalign{\smallskip}
            \hline
         \end{tabular}
\begin{itemize}
\item[$^a$] The projected semi-major axis $a \sin i$ is the
  semi-major axis of the projection of the orbit of the pulsar, around
  the system's center of mass, onto the plane containing the line of
  sight and the line of nodes.
\item[$^b$] Characteristic age and surface magnetic field have been
  calculated using standard formulas, namely $\tau_{\rm
  c}=P/2\dot{P}$ and
  $B_{0}=3.2\times10^{19}\sqrt{P\dot{P}}$\,G. The total mass $M_{\rm
  TOT}$ has been calculated from the relativistic
  periastron advance and the measured Keplerian parameters,
  assuming the validity of general relativity.
  The minimum companion mass was estimated
  using the observed mass function $f(M_{\rm
  C})$ and the lower limit for the total mass, as given by its
  uncertainty, in the case of $\sin i = 1$. For details see
  Lorimer \& Kramer (2005)\nocite{lk05}.
\end{itemize}

\end{table}

\begin{figure}
\centering
\includegraphics[angle=0,width=8.5cm]{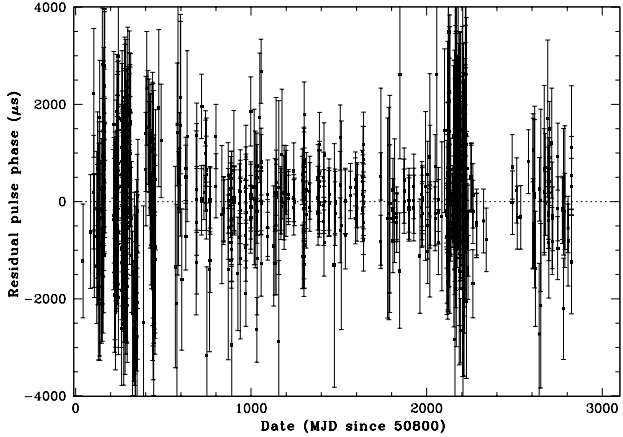}
\caption{Timing residuals after jointly applying the final model to
all three data sets. Vertical bars represent the ToA's uncertainty.}
\label{fig:jointres}
\end{figure}

\section{The nature of the companion}

In their discovery paper, \citet{lcm+00} proposed that this system is
a member of the small class of DNS binaries. Soon after, \citet{mig00}
reported on optical observations of the region surrounding the pulsar
position to search for emission from the pulsar companion. They
detected no emission coincident with the pulsar position and while not
conclusive, the lack of emission is at least consistent with the
neutron star hypothesis for the nature of the companion.

The derived values for the characteristic age ($\tau_c = 1.83 \times
10^{9}$ yrs) and the surface magnetic field ($B = 9.8 \times
10^{9}$G), as well as the combined values of the spin period
($P=104$~ms) and its period derivative ($\dot{P}=9\times10^{-19}$),
indicate that PSR\,J1811$-$1736 is a neutron star that experienced a
spin-up phase via accretion from mass overflowing from its
companion. These parameters, in conjunction with the measured orbital
eccentricity ($e=0.828$), indeed suggest that PSR\,J1811$-$1736 is a
mildly recycled pulsar whose companion star was massive enough to also
undergo a SN explosion. This second SN imparted the actually observed
large eccentricity to the system (e.g. \citealt{bv91}).

Our new measurement of the relativistic periastron advance,
$\dot{\omega} = 0.0090 \pm 0.0002$ deg yr$^{-1}$, allows us to
determine the value of 2.57$\pm$0.10 M$_{\odot}$ for the total mass of
the system, assuming that general relativity is the correct theory of
gravity and that the observed value is fully due to relativistic
effects (e.g. \citealt{dd86}).  This value, combined with the measured
mass function, implies a minimum companion mass of 0.93 $M_{\odot}$.

\begin{figure}
\centering \includegraphics[angle=0,width=8.5cm]{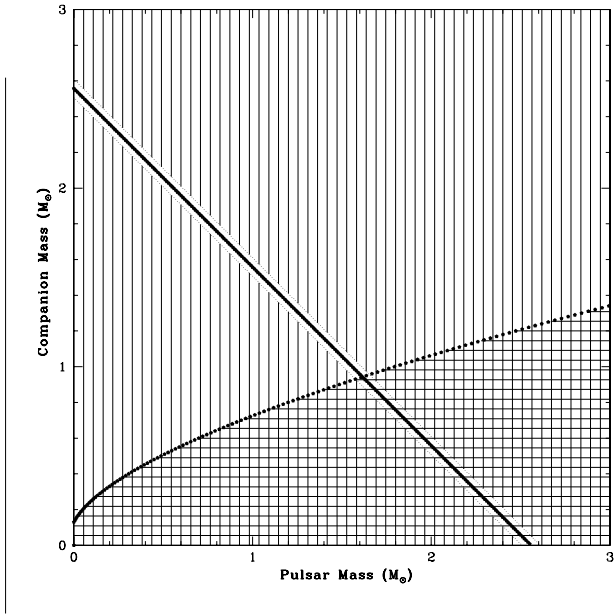}
\caption{Mass-mass diagram for the binary system hosting
PSR\,J1811-1736. The shaded area below the dotted curved line is
excluded because of the geometrical constraint $\sin i \leq 1$, while
the area outside the diagonal stripe is excluded by the measurement of
the relativistic periastron advance and the derived value for the
total mass for this system.}
\label{fig:masses}
\end{figure}

The value for the total mass is relatively low, but very similar to
the total mass of the double pulsar system \citep{lbk+04} and the
recently discovered DNS system PSR\,J1756$-$2251 \citep{fkl+05}. In
fact, these systems have neutron star companions that have the lowest
neutron star masses observed so far, $M_{c}~=~1.25M_{\odot}$ and
$M_{c}=1.17M_{\odot}$ (for a recent review see 
Stairs 2004\nocite{sta04a}), respectively.  In Figure \ref{fig:masses} we
show the so-called mass-mass diagram where the pulsar and companion
masses can be directly compared. The measured value for the advance of
periastron means that the sum of the masses must lie along the
diagonal line, while the constraint on the inclination $\sin i \leq 1$
excludes the hatched region below the dotted line. Assuming that the
neutron stars in this system must have a mass which is larger than the
{\it lowest} mass so-far measured, i.e. $1.17M_\odot$, we find that
they both have masses in the interval
1.17$\,M_{\odot}\,\leq\,M_{P},M_{C}\,\leq\,1.50 M_{\odot}$. This
interval contains all but the heaviest neutron stars masses for which
a reliable determination has been obtained. Using this mass
constraint, we can also translate this range into lower and upper
limits on the inclination of the system, i.e. 44\,deg\,$\lapp i
\lapp$\,50\,deg.

Alternatively, if either the pulsar or the companion have a mass equal
to the observed median neutron star mass of 1.35 M$_{\odot}$
\citep{sta04a}, the other neutron star would have a mass of
$1.22\,\pm\,0.10 M_{\odot}$. This value is consistent with the lower
limit in the previous discussion, but this also allows for the
possibility that one of the two neutron stars has a mass as low as
1.12 $M_{\odot}$.

\section{Future potential of timing observations}

In order to determine the companion mass without ambiguities, it is
necessary to measure a second post-Keplerian (PK) parameter
(e.g. \citealt{dt92}). We have investigated the possibility of
measuring the PK parameter $\gamma$, which describes the combined
effect of gravitational redshift and a second order Doppler effect. For
a companion mass of $1.35M_\odot$, the expected value is $\gamma =
0.021$ ms. Using simulated data sets for the presently available
timing precision, we estimate that a $3\sigma$ detection for $\gamma$
is achievable after about 4 more years of observation. However, in
order to obtain a 10\% accuracy in mass measurement by determining
$\gamma$ to a similar precision, several decades of observations may
be needed.

From similar simulations, we estimate that PK parameters like the rate
of orbital decay, $\dot{P}_{\rm B}$, or the Shapiro delay, are
unmeasurable in this system, unless significant improvements in timing
precision can be obtained.  For a companion mass of 1.35 $M_{\odot}$,
$\dot{P}_{\rm B}$ is only $-9.4 \times 10^{-15}$~s~s$^{-1}$ and we
expect an amplitude of only 6\,$\mu$s for the amplitude of the Shapiro
delay. We also note that the effects of geodetic precession (see
e.g. \citealt{kra98}) will not be measurable within a reasonable time,
as it has a period of order of $10^{5}$ years.

\section{Improving timing precision}

It is obvious that the measurement of further PK parameters will only
be possible if higher timing precision can be achieved for this
pulsar. For instance, if a precision of 50$\mu$s could be obtained,
$\gamma$ could be measurable to a 10\% accuracy after a total of just
5 yr of observations, while a $3\sigma$ detection of the orbital decay
may be achieved after about 6 yr.

One way to achieve higher timing precision is to detect narrow
features in the observed pulse profile by means of higher effective
time resolution.  This is most commonly achieved through better
correction for dispersion smearing that is  caused by the radio
signal's passage through the ionized  interstellar medium.  While this
effect is actually completely removed by the  use coherent
de-dispersion techniques (see \citealt{hr75}) at some of our
telescopes, it is apparent that the current timing precision is limited
instead by broadening of the pulse profile due to interstellar
scattering (\citealt{lmg+04}, and references therein). Indeed, the
pulse profile at 1.4~GHz shows a strong scattering tail (see Fig.\,3)
which prevents a highly accurate determination of the pulse time of
arrival. As scattering is a strong function of observing frequency, we
can expect to reduce its effect, and hence to enable higher timing
precision, by using timing observations at frequencies above  1.4~GHz.

\begin{figure}
\centering \includegraphics[angle=0,width=8.5cm]{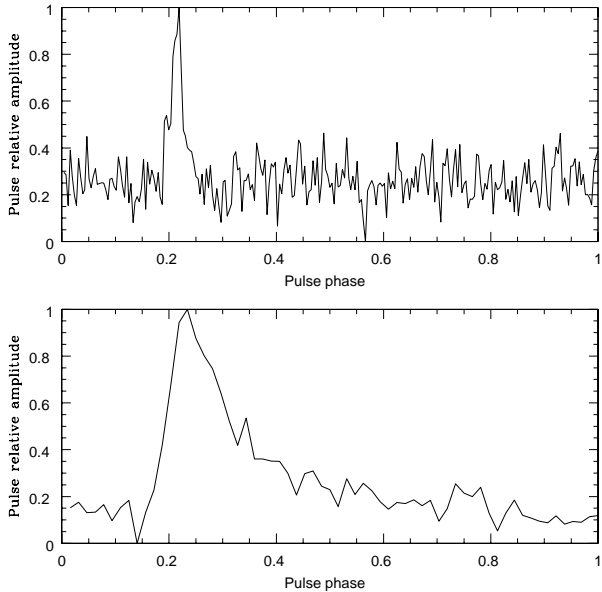}
\caption{Pulses' profiles of PSR\,J1811--1736 at 1.4\,GHz (bottom
panel) and 3.1\,GHz (top panel). Both profiles have been obtained with
10 minutes observations performed with the Parkes radio telescope in
February 2005.}
\label{fig:3GHzp}
\end{figure}

We obtained observations at 3.1~GHz that confirm this expectation.
The pulse profile obtained  at this frequency shows no evidence of
interstellar scattering, and its width at 10\% is only 7.3~ms. This is
a great improvement with respect to the 1.4~GHz profile, whose 10\%
width is 58.3~ms. A flux density of $0.34\pm0.07$mJy  measured at
3.1\,GHz suggests that regular timing observations at this frequency
should be possible and should significantly improve the achievable
timing precision. This would allow us to measure a second PK parameter
to an accuracy that is sufficiently precise to determine the companion
mass.

Using the data available at 1.4\,GHz and 3.1\,GHz, we computed
spectral indexes for flux density and scattering time.  For the flux
density we obtain a spectral index of $\beta=-1.8\pm0.6$.  Subdividing
our observing band at 1.4\,GHz we obtain two different profiles that
we use to measure the pulse scatter timescale $\tau$ by applying the
technique described in \citealt{lkm+01}. We convolve  the
3.1\,GHz-profile, assumed to represent the true pulse shape, with an
exponential scattering tail and obtain scattering times by a
least-square comparison of the convolved profile with the observed
pulse shape. At 1.284\,GHz we find $\tau_s~=~16.9$~ms, and
$\tau_s~=~10.6$~ms at 1.464~GHz, respectively.  This results in a
spectral index $\alpha$ of the scattering time, i.e.~$\tau \propto
\nu^{-\alpha}$, of $\alpha~=~3.5\pm~0.1$. Such a measured value agrees
very well with analogous results from L\"{o}hmer et
al. (\citeyear{lkm+01,lmg+04}) who determined $\alpha$ for a number of
pulsars with very high dispersion measures.

The measured spectral index of the scattering time is also consistent
with the fact that the pulsar has not been detected at frequencies
below 1\,GHz. For example at 400\,MHz we calculate $\tau_{s}\sim1$\,s
which is almost an order of magnitude greater than the spin period of
the pulsar thus making it impossible to detect it as a pulsating source.

\section{Previous searches for pulsations from the companion}

Searches for pulsations from the binary companion of PSR\,J1811-1736
have been performed on Parkes and Effelsberg data. Parkes observations
have been investigated with the procedure described in \citet{fsk+04},
while Effelsberg data have been processed using the procedure
described in \citet{kle04}. Both searches were unsuccessful in
detecting any evidence of pulsation. The very high value of the
dispersion measure for this system may suggest that the interstellar
scattering is responsible for the failure in detecting any
pulsation. Therefore we studied the possible impact of this phenomenon
on our searches for pulsations from the companion of PSR\,J1811-1736.

We considered 1\,hr observations done with the Effelsberg telescope
using either the 20\,cm (1.4\,GHz) or the 11\,cm (2.7\,GHz) receiver,
exploring a range of possible flux densities ($S=0.05, 0.5, 1$\,mJy)
and a detection in a signal-to-noise ratio threshold of
$S/N=10$. Using the DM of the observed pulsar, and assuming Effelsberg
observations at 1.4\,GHz, we find a minimum detectable period of
$P_{\rm min}=750$\,ms for a flux density of $S=50\,\mu$Jy, while even
for the flux densities of $S=500\,\mu$Jy and $S=1\,$mJy periods below
$\sim$10\,ms become undetectable at the observing frequency of
1.4\,GHz.

At 2.7\,GHz, the system performance of the Effelsberg telescope allows
for an antenna gain of $G=$1.5\,K\,Jy$^{-1}$ with a system temperature
$T_{\rm sys}=17$\,K. Using these parameters, we obtain minimum periods
of $P_{\rm min}=110$\,ms, $P_{\rm min}=2.5$\,ms and $P_{\rm min}=
1.6$\,ms  for flux densities $S_{\rm 2.7\,GHz}=50\,\mu$Jy,
500\,$\mu$Jy and 1\,mJy respectively.

When searching for pulsations from the binary companion of a Galactic
recycled pulsar in a double neutron star system, it is more likely
that the companion is a young pulsar with rather ordinary spin
parameters, as found for  PSR\,J0737-3039B in the double pulsar system
\citep{lbk+04}. Our lower limits on the minimum detectable period
therefore suggests that  interstellar scattering should not have
prevented the detection of the companion, unless it were a very fast
spinning or very faint source.

\section{Constraints on the kick velocity of the second SN explosion}

The large eccentricity of J1811$-$1736  system can be ascribed to a
sudden loss of mass which results in a change of the orbital
parameters. Such a sudden loss of mass can be attributed to the SN
explosion that formed the younger unseen neutron star companion  (see,
e.g., \citealt{bv91}). Under the hypothesis of a symmetric explosion,
simple calculations show that the binary survives this event only if
the expelled mass $M_{\rm exp}$ is less than half of the total mass of
the binary before the explosion (pre-SN binary). The induced
eccentricity is a simple function of the amount of the expelled mass:
$e=M_{\rm exp}/M_{\rm TOT}$, where $M_{\rm TOT}$ is the total mass of
the pre-SN binary. In the case of the binary system hosting
PSR\,J1811-1736, the measured eccentricity, $e=0.828$, and the derived
total mass, $M_{\rm bin}=2.57\,M_{\odot}$, would imply a total mass
$M_{\rm TOT}\,=\,4.7\,M_{\odot}$ for the pre-SN binary.

The high space velocities measured for isolated pulsars indicate that
neutron stars may receive a kick when formed, with an unpredictable
amplitude and direction \citep{hp97,cc98,acc02}. Such kicks
imparted to the newly formed neutron stars are caused by {\em
asymmetric supernova explosions}. If an asymmetric SN explosion
occurs in a binary system, the survival and the eventual post-SN
binary parameters are jointly determined by the mass loss and the
vector representing the velocity imparted to the neutron star. In this
case a simple survival condition like the one derived for the
symmetric explosion case cannot be determined.

A correlation between the pulsar's spin period and orbital
eccentricity has recently been found for DNS systems
(\citealt{mlc+05}, \citealt{fkl+05}). A numerical simulation by
\citet{dpp05} linked this correlation to the typical amplitude of the
kick velocity received by the younger neutron star at birth.
\citet{dpp05} found that the spin period versus eccentricity
correlation is recovered if the typical kick amplitude satisfies the
condition $V_{\rm K} \lapp 50$\,km\,s$^{-1}$.

To investigate the nature of the kick received by the younger neutron
star in this system we considered as the pre-SN binary a binary system
containing a neutron star and a helium star. We then constrained the
total mass of this system by combining our results on the total mass
of the actual binary with the range given by \citet{dp03b} for the
mass of the helium star that was the companion of PSR\,J1811-1736
before the explosion. The helium star mass range $2.8M_{\odot} \leq
M_{\rm C} \leq 5.0 M_{\odot}$ given by \citet{dp03b} leads to a total
mass range of $4.0M_{\odot} \leq M_{{\rm {TOT}}} \leq 6.5M_{\odot}$.

Binary parameters for the pre-SN binary have been chosen as follows.
The eccentricity has been assumed negligible, since the accretion
phase responsible for spinning up the pulsar also provided strong
tidal forces that circularised the orbit. The orbital separation has
been constrained to be between the the minimum (pericentric) and
maximum (apocentric) distance between the two neutron stars in the
post-SN binary. This statement can be justified as follows. The
typical velocity of the expelled matter in a SN explosion is close to
the speed of light, while the typical orbital velocity of the stars in
a binary system is of order of $\sim 100$\,km\,s$^{-1}$, using for the
total mass any value in the range we used for $M_{{\rm {TOT}}}$ and a
value for the orbital separation comparable to the one for the present
binary system, i.e. few light-minutes (see the discussion in the next
paragraph of the post-SN binary evolution due to general relativistic
effects). This means that the change in position of the two stars is
negligible if compared to the change in position of the expelled
matter. The time required for the binary system to do the transition
from the pre-SN to the post-SN binary is the time required by the
expelled matter to travel along a path as long as the orbital
separation, i.e. few minutes. After such an elapsed time the matter
expelled in the SN explosion encloses both stars and has no more
gravitational effects on their binary motion. This time is also much
shorter than the orbital period of few days for a pre-SN binary like
the one we are considering. This means that during this transition the
positions of the two stars remained unchanged, and their distance was
a distance periodically assumed by the two stars also in their orbital
motion in the post-SN binary.

To make a fully consistent comparison between the actually observed
binary and the eccentric binary that emerged from the last SN
explosion (post-SN binary) one has to take into account the secular
changes of the orbital parameters caused by general relativistic
effects.  In order to do this one needs to have an estimate for the
time since the last SN. The only timescale that is available to us is
the characteristic age of the first born pulsar. We then find binary
parameters that are consistent with the present system values within
their uncertainties. Given the well known uncertainties in this age
estimation, we considered the possibility that the present binary is
in fact up to ten times older than suggested by the characteristic age
(i.e 1.8$\times 10^{10}$ yr). Even when considering this extreme age,
we find that our results remain unaffected.  We consequently decided
to use as post-SN orbital parameters the same values we measure today.

\begin{figure}
\centering
\includegraphics[angle=0,width=8.5cm]{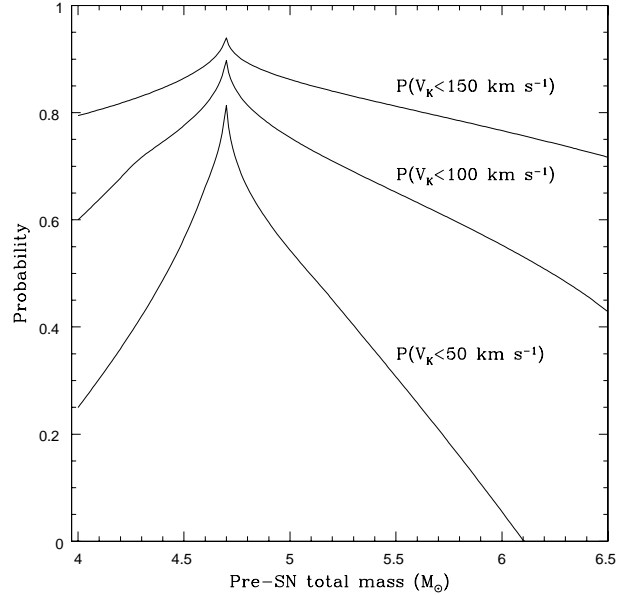}
\caption{Probabilities to have a kick velocity lower than or equal to
50 (lower line), 100 and 150 (upper line) km s$^{-1}$ as a function of
the pre-SN total mass. All probabilities peak in correspondence of a
pre-SN total mass $M_{\rm TOT}=4.70M_{\odot}$, which is the mass of
the binary system before the explosion in the case of a symmetric
SN. The probability to have a kick velocity lower than 50 km
s$^{-1}$ is not negligible for all but the highest considered values
for the binary pre-SN mass.}
\label{fig:probs}
\end{figure}

By insisting that the total energy and total angular momentum,
calculated in the center-of-mass frame, of the post-SN binary and the
present systems are conserved, we can combine these terms to obtain an
equation for the kick amplitude, as a function of the two angles,
representing its direction in a suitable reference frame, and the
total mass and orbital separation before the explosion. We assumed
that the probability of occurrence for any given kick vector is
proportional to the solid angle described by the direction of the kick
in spherical coordinates and then calculated the probability of having
a kick velocity lower than some fixed values. We chose the values of
50, 100 and 150~km~s$^{-1}$ ( hereafter $P_{50}$, $P_{100}$ and
$P_{150}$ respectively.).  Figure\,\ref{fig:probs} shows that $P_{50}$
is not negligible if the total mass of the pre-SN binary system is
lower than 6\,$M_{\odot}$. Moreover, all considered probabilities peak
in correspondence of a total pre-SN mass of 4.70\,$M_{\odot}$,
corresponding to the null kick case. These results lead to the
conclusion that the younger neutron star in this system received a low
velocity kick and is thus similar to all other known DNS systems,
which all have tighter orbits.

Nevertheless,  the binary system containing PSR\,J1811-1736 is much
wider than all other known DNS systems. This may indicate that the
binary evolution of this system may have been (at least partially)
different. In particular the wide  orbital separation for this system
may be compatible with an evolution during which the pulsar's
progenitor avoided completely a common envelope phase \citep{dp03b} or
that this phase was too short to sufficiently reduce the orbital
separation. Moreover if the spin-up occurred via the stellar wind of
the giant companion then the system would tend to be wider due to the
isotropic mass loss from the companion \citep{dpp05}.

\section{Summary \& Conclusions}

We have presented an improved timing solution for the binary pulsar
J1811$-$1736. This solution improves the previously measured values
for the spin and Keplerian orbital parameters and one post-Keplerian
orbital parameter, the periastron advance. These results would not
have been achieved without data from the three telescopes used and are
the first obtained as part of the European Pulsar Timing Array (EPTA)
collaboration.

The measured values for the spin period and its first derivative are
typical of a mildly recycled neutron star, while the high eccentricity
of the binary system can be seen as a signature of the SN explosion
that interrupted the mass transfer from the companion to the accreting
neutron star. This is likely to have occurred before the pulsar could
reach spin periods typical of the fully recycled (i.e. millisecond)
pulsars. This leads to the conclusion that PSR\,J1811$-$1736 is a
member of a DNS binary system.

The determined  value of the periastron advance provides further
confirmation for this scenario as it suggests a total mass of the
system of $M_{tot}$~=~2.57$\pm$0.10~$M_{\odot}$.  This value is
similar to the total mass of two other DNS systems, i.e. the double
pulsar \citep{lbk+04} and PSR\,J1756-2251 \citep{fkl+05}. In both
these systems, the non-recycled neutron stars is very light. Assuming
that PSR\,J1811$-$1736 is a neutron star with a mass within the
currently measured mass range for neutron stars, we find the companion
mass to lie in the same range. Using these arguments we determine the
inclination of the orbital plane to within $\sim6$ degrees.

We also investigated the possibility of measuring a second
post-Keplerian parameter, in order to determine both masses and thus
to definitively determine the nature of the companion. Unfortunately, the
pulse profile at 1.4~GHz is heavily broadened by interstellar
scattering which limits the timing precision and means that a second
post-Keplerian parameter is not measurable within a reasonable amount
of time with observations at that frequency.  However we find that at
3~GHz the scattering is sufficiently reduced and the flux density is
sufficiently high that higher precision timing will be possible at
this frequency. Comparing the pulse profiles at 1.4 and 3~GHz we find
that the scattering timescale for this pulsar scales with frequency
with a power law of index $\alpha~=~3.5\pm0.1$ which is in excellent
agreement with earlier results on high dispersion measure pulsars.
  
Considering the effects of the interstellar scattering on the
detectability of pulsations from the companion, we find that the
minimum detectable period is longer at lower frequencies and for
fainter objects. In general, we do not expect interstellar scattering
to be the cause for the continued non-detection of the companion
neutron star.

The orbital separation for this system is much wider than for all
other DNS systems, and it suggests that its binary evolution has been
different. One explanation invokes the lack of a common envelope
phase, during which the size of the orbit shrinks due to tidal forces
in the envelope of the companion star. Another explanation
\citep{dpp05}, not necessarily conflicting with the previous one,
invokes a different mass transfer mechanism in the spin-up phase of
the pulsar, namely via stellar wind, while all other recycled pulsars
in the known DNS have been spun up via Roche lobe overflow mass
transfer.

Finally, we investigated the kick imparted to the second born neutron
star during the second SN.  We find that for realistic values of the
total mass of the pre-SN binary, the kick velocity has a not
negligible probability of being lower than 50~km~s$^{-1}$. This
constraint is common to all DNS systems, as shown by
\citet{dpp05}. This evidence for a low amplitude asymmetric kick
received by the younger neutron star may be the consequence of the
effects of binary evolution on a star that undergoes a SN explosion,
effects that somehow are able to tune the amplitude of such kick.

\end{document}